\documentclass[useAMS,usenatbib]{mn2e}
\usepackage{times}
\usepackage{graphicx}
\usepackage{amsmath}
\usepackage[T1]{fontenc}
\usepackage{aecompl}

\begin{document}

\title[DCBH formation from synchronized halos]{Direct collapse black hole formation from synchronized pairs of atomic cooling halos}

\author[E. Visbal et al.]{Eli Visbal$^1$\thanks{visbal@astro.columbia.edu} \thanks{Columbia Prize Postdoctoral Fellow in the Natural Sciences}, Zolt\'{a}n Haiman$^1$, Greg L. Bryan$^1$ \\ $^1$Department of Astronomy, Columbia University, 550 West 120th Street, New York, NY, 10027, U.S.A. }

\maketitle

\begin{abstract}
High-redshift quasar observations imply that supermassive black holes (SMBHs) larger than $\sim 10^9 ~ M_\odot$ formed before $z \sim 6$.  That such large SMBHs formed so early in the history of the Universe remains an open theoretical problem.  One possibility is that gas in atomic cooling halos exposed to strong Lyman-Werner (LW) radiation forms $10^4-10^6 ~ M_\odot$  supermassive stars which quickly collapse into black holes.  We propose a scenario for direct collapse black hole (DCBH) formation based on synchronized pairs of pristine atomic cooling halos. We consider halos at very small separation with one halo being a subhalo of the other. The first halo to surpass the atomic cooling threshold forms stars.  Soon after these stars are formed, the other halo reaches the cooling threshold and due to its small distance from the newly formed galaxy, is exposed to the critical LW intensity required to form a DCBH.  The main advantage of this scenario is that synchronization can potentially prevent photoevaporation and metal pollution in DCBH-forming halos. We use N-body simulations and an analytic approximation to estimate the abundance of DCBHs formed in this way. The density of DCBHs formed in this scenario could explain the SMBHs implied by $z\sim 6$ quasar observations.  Metal pollution and photoevaporation could potentially reduce the abundance of DCBHs below that required to explain the observations in other models that rely on a high LW flux.
\end{abstract}

\begin{keywords}
quasars: supermassive black holes--cosmology: theory
\end{keywords}

%\onecolumn
\section{Introduction}
High-redshift quasar observations imply that supermassive black holes (SMBHs) larger than  $ \sim 10^9 M_\odot$ formed by $z=6$.  How these black holes grew so large, so quickly, remains an open theoretical problem \citep[see reviews by][]{2013ASSL..396..293H,2010A&ARv..18..279V}.  Black hole seeds produced by the first metal free (pop III) stars would need to accrete at the Eddington limit for the entire age of the Universe in order to grow to  $3 \times 10^9 {\rm M_\odot}$ by $z=6$ (assuming $10$ per cent radiative efficiency and an initial mass of $100 ~ M_\odot$).  However, radiative feedback is expected to prevent uninterrupted Eddington limited accretion over such a long period of time \citep{2007MNRAS.374.1557J,2009ApJ...701L.133A,2009ApJ...698..766M}.

A possible solution to this problem is that SMBH seeds form through direct collapse of gas in atomic cooling ($T_{\rm vir} \ga 10^4 \rm{K}$) halos, resulting in $10^4-10^6 M_\odot$ supermassive stars or quasi-stellar envelopes which quickly collapse into black holes (for reviews see \citealt{2013ASSL..396..293H,2010A&ARv..18..279V}).  The larger mass of these black hole seeds compared to pop III remnants reduces the tension between the formation time of $z=6$ quasars and the age of the Universe.  However, these black holes must still accrete gas over a significant fraction of the available time \citep{2009ApJ...696.1798T, 2014arXiv1406.3023T}.  Another possibility is that black hole seeds accrete faster than the Eddington limit at high redshift \citep[][]{2014ApJ...784L..38M}, but we do not consider that scenario in this work.  

To form a DCBH, efficient molecular hydrogen cooling leading to gas fragmentation and star formation must be prevented.  This most likely requires a strong Lyman-Werner (LW) background, which destroys molecular hydrogen \citep{2014arXiv1403.1293V}.  Simulations and one-zone models have shown that the critical LW intensity required to form a DCBH is, depending on the spectrum, $J_{\rm crit} \sim 1000$ (in units of $10^{-21} \rm{ergs/s/cm^2/Sr/Hz}$) \citep[][]{2010MNRAS.402.1249S, 2011MNRAS.418..838W,2012MNRAS.425L..51W, 2014arXiv1404.5773L} .  \footnote{There has been much confusion in the literature regarding the proper $J_{\rm crit}$ to use.  Many authors have used $J_{\rm crit}=30$ for pop II LW sources, corresponding to a $T=10^4~ \rm{K}$ blackbody spectrum.  However, a realistic pop II spectrum will not be a $T=10^4~ \rm{K}$ blackbody.  For example, assuming 10 Myr of continuous star formation with a Salpeter IMF and a metallicity of $Z=0.001$, the population synthesis models of Starburst99 \citep{1999ApJS..123....3L} predict spectra more similar to blackbodies with $T \sim \rm{several} \times 10^4 ~\rm{K}$ than $T \sim10^4~ \rm{K}$.  For blackbody spectra with $T=2 \times 10^4 - 10^5 ~ \rm{K}$, one-zone models predict $J_{\rm crit} \sim 1000$ (for more details see Wolcott-Green et al., in prep).}

The critical LW intensity is much higher than the expected mean background \citep[see e.g.][]{2009ApJ...695.1430A, 2013MNRAS.432.2909F,2014arXiv1402.0882V}.  Thus, in order for an atomic cooling halo to be exposed to $J_{\rm crit}$, it must have a bright nearby galaxy.  Analytical and numerical studies have aimed to quantify the abundance of close pairs of halos which could potentially achieve $J_{\rm crit}$ \citep{2008MNRAS.391.1961D,2014arXiv1405.6743D,2012MNRAS.425.2854A,2014arXiv1403.5267A,2014MNRAS.440.1263Y} and have found that there may be enough close pairs to explain the abundance of bright $z=6$ quasars.  

However, one potential obstacle to DCBH formation that requires more careful study is whether the progenitors of an atomic cooling halo will be ionized and photoevaporated before a DCBH is formed.  This could prevent DCBH formation because the atomic cooling halo will lack a dense gas core.  Even if, after the halo reaches the atomic cooling threshold, gas falls back in, the higher electron fraction could catalyze the formation of molecular hydrogen effectively raising $J_{\rm crit}$ \citep{2014arXiv1405.2081J}.  

Recently, \cite{2014arXiv1405.6743D} estimated the abundance of DCBHs taking into account supernovae winds from neighboring halos that pollute the intergalactic medium (IGM) with metals and prevent DCBH formation.  In these models, DCBHs form near galaxies in massive, $\sim 10^{11}-10^{12} M_\odot$ halos, at distances of $\sim 30 ~ \rm{kpc}$.  However, we argue that the progenitors of DCBH-forming halos near such large galaxies will be susceptible to early photoevaporation due to the following argument.  A typical $10^{12} M_\odot$ halo at $z=10$ had one or more $\sim 10^{11} M_\odot$ progenitors at $z=15$ (which we determine with the hybrid progenitor mass function from \cite{2004ApJ...609..474B}).  Assuming $\sim 10$ ionizing photons released into the IGM for each baryon collapsing into a dark matter halo over 10 percent of the Hubble time and an IGM clumpiness of $C = \langle n_{\rm H}^2 \rangle/\langle n_{\rm H} \rangle^2 \sim 3$  \citep{2009MNRAS.394.1812P,2012MNRAS.427.2464F}, a $ 10^{11} M_\odot$ halo will quickly (in $<10 ~ \rm{Myr})$ produce an ionized bubble with $r \sim 65 ~\rm{kpc}$.    
Utilizing the fits of \cite{2005MNRAS.361..405I} (assuming a $T=5\times10^4 ~\rm{K}$ blackbody spectra) to determine the photoevaporation time and the spherical collapse model to estimate the maximum distance between the DCBH-forming halo's progenitors and the $ 10^{11} M_\odot$ halo at $z\sim15$, we find that the progenitors of the DCBH-forming halo can be photoevaporated in less than $200~\rm{Myr}$.    The time from $z=10-15$ corresponds to $\sim 200 ~\rm{Myr}$, so by $z=10$ the DCBH-forming halo is likely to be ionized, effectively raising $J_{\rm crit}$.

\cite{2014arXiv1405.6743D} assume $J_{\rm crit} = 300$ in their fiducial model and find that the number density of DCBH is roughly $\sim 100$ times larger than that necessary to explain the brightest $z=6$ quasars.  However, if $J_{\rm crit} \sim 1000$ or higher and the effect of photoevaporation is important, the abundance they compute may be too low to explain observations.

In this paper, we introduce a DCBH formation scenario that bypasses the obstacle of photoevaporation and potentially the issue of metal contamination from supernovae winds.  We consider a pair of halos at very small separation that cross the atomic cooling threshold at nearly the same time. Molecular cooling and star formation is likely to be prevented in these halos' progenitors due to the mean cosmic LW (this is discussed in detail in \S4).  The first halo in the pair to cross the cooling threshold forms stars.  The halos merge, with the second halo orbiting the first as a subhalo.  At nearly the same time stars are made in the first halo, the second halo crosses the atomic cooling threshold.  Because it is very close to the newly formed galaxy in the first halo, it is exposed to $J_{\rm crit}$ and forms a DCBH.  This scenario avoids photoevaporation because, provided the two halos are not in a large-scale ionized region (most of the IGM is expected to be neutral at $z>10$), the DCBH-forming subhalo is only exposed to ionizing radiation for a short period of time.  This picture may also avoid metal pollution because, depending on the mass of the metal free stars formed in the star-forming halo, the DCBH can form before the stars reach the end of their lives and expel metals through supernovae winds.

This paper is structured as follows. In \S2, we summarize the synchronized halo scenario and show that ram pressure stripping, tidal disruption, photoevaporation, and metal contamination are unlikely to prevent DCBH formation.  We predict the abundance of DCBHs formed this way in \S3 using both analytical calculations and N-body simulations.  In \S4 we discuss how regions large enough to host the brightest $z=6$ quasars will form DCBHs at higher redshifts than our numerical simulations, but that there will still be a high enough number density of DCBHs to explain the observations.   We also compute how high the LW background must be to prevent star formation and subsequent metal pollution in the progenitors of atomic cooling halos. We summarize our results and discuss our conclusions in \S5.  Throughout we assume a $\Lambda$CDM cosmology consistent with the latest constraints from Planck \citep{2013arXiv1303.5076P}: $\Omega_\Lambda=0.68$, $\Omega_{\rm m}=0.32$, $\Omega_{\rm b}=0.049$, $h=0.67$, $\sigma_8=0.83$, and $n_{\rm s} = 0.96$.  Throughout we use a "c" to distinguish between comoving and physical units.

\section{DCBH formation through synchronized atomic cooling halos}
As described above, we consider DCBH formation through the synchronized formation of two atomic cooling halos at small separation.  In particular, we consider halos close enough for one to become a subhalo of the other.   The first halo forms a galaxy as the second halo crosses the atomic cooling threshold.  The galaxy provides enough LW radiation to shut off molecular cooling leading to DCBH formation.   In this section, we estimate the time it takes the core of an atomic cooling halo to collapse and the flux required from the galaxy-forming halo to produce a DCBH.  We also show that ram pressure stripping, tidal disruption, and photoevaporation should not pose serious problems.  Finally, we discuss how metal enrichment in the DCBH-forming subhalo from supernovae winds could be avoided in this picture.

\subsection{Collapse time}
Here we compute how long it takes the gas core of a $T_{\rm vir}=10^4 \rm{K}$ atomic cooling halo to collapse in free fall. The corresponding virial mass is $M_{\rm c} = 3 \times 10^7 M_\odot$ at $z=10$ assuming a mean molecular weight of $\mu=1.22$ for neutral primordial gas. We assume a static NFW dark matter profile with $c=5$ \citep{1997ApJ...490..493N} and a constant density baryon core with density $n_{\rm core} = 5 ~ \rm{cm}^{-3}$ and radius of $r_{\rm vir}/10$.  This is consistent with cosmological simulations without radiative cooling \citep{2014arXiv1403.1293V}.  Solving numerically for the trajectory of the edge of the gas core, we find that it takes approximately $t_{\rm coll} \sim 10 ~ \rm{Myr}$ for complete collapse of the core.  Throughout this paper we use this as the time it takes a halo to create stars or a DCBH after it crosses the atomic cooling threshold.  Due to the self-similar nature of gas and dark matter density profiles we expect this characteristic time to scale with redshift as $t_{\rm coll} \propto (1+z)^{-3/2}$.  Note that the cooling timescale, $t_{\rm cool} = 1.5 k_{\rm B} T n_{\rm gas}/(n_{\rm H}^2 \Lambda)$,  in an atomic cooling halo is expected to be much smaller than $t_{\rm coll}$.

\subsection{Required flux}
In order for the second halo to form a DCBH, we assume it must be exposed to $J_{\rm crit}\sim 1000$ (Wolcott-Green et al., in prep).  We assume this halo must be exposed to $J_{\rm crit}$ for the entire duration of its collapse  ($t_{\rm coll} =10 \rm{Myr}$).  This is a conservative assumption, since a weaker flux will be required to suppress molecular cooling early on when the density is low. 

 To maintain $J_{\rm crit}$ on the DCBH-forming halo, a star-forming halo at separation $d$ must have a LW luminosity of
\begin{equation}
N_{\rm LW, crit} = 1.1 \times 10^{52} \left ( \frac{d}{0.5~ \rm{kpc}} \right )^2 \left ( \frac{J_{\rm crit}}{1000} \right ) \rm{photons/s}.
\end{equation}
We consider a star-forming galaxy formed from metal free gas.  To estimate the LW flux, we use the lifetime averaged stellar properties computed in \cite{2002A&A...382...28S}.  Assuming $d=0.5~ \rm{kpc}$ and $J_{\rm crit}=1000$, if all of the stars are $5 M_\odot$, $19$ per cent of the gas in the halo must form stars to achieve $1.1 \times 10^{52}~ \rm{LW~ photons/s}$.  More massive stars produce more LW photons per baryon and would require less efficient star formation.  For the same $d$ and $J_{\rm crit}$, if all stars were $15 M_\odot$ only $1.4$ per cent of the gas would need to turn into stars to achieve this flux ($0.18$ per cent for $120 M_\odot$).  Given the large uncertainty associated with the pop III initial mass function (IMF) in an atomic cooling halo, it is impossible to precisely predict the flux from the first halo.  However, even if a majority of the stars are small ($\sim 5 M_\odot$) it would still be plausible to achieve $J_{\rm LW} =1000$ at close distances ($\sim 4.75$ per cent of the gas would have to form stars for $d=0.25 ~\rm{kpc}$).

\subsection{Ram pressure stripping and tidal disruption}
Next, we consider whether the cores of atomic cooling halos will be disrupted by ram pressure stripping as they orbit one another at small distances.  Following the analytic treatment of \citep{2008MNRAS.383..593M}, the core of a halo will remain intact provided
\begin{equation}
\label{ram}
P_{\rm ram}  = \rho_{\rm m} v^2 \le  \frac{\alpha G M_{\rm tot}(R_{\rm core}) \rho_{\rm gas}(R_{\rm core})}{R_{\rm core}},
\end{equation}
where $\rho_{\rm m}$ is the gas density the core passes through, $v$ is the relative orbital velocity, and $\alpha$ is an order unity parameter set by the gas and dark matter density profiles.  $M_{\rm tot}(r)$ and $\rho_{\rm gas}(r)$ are the total mass (gas plus dark matter) contained within radius $r$ and the gas density at $r$, respectively.  As before, we assume an NFW dark matter profile with $c=5$ and a gas profile with a constant core out to $ r_{\rm vir}/10$ and a $\rho \propto r_{\rm gas}^{-2}$ envelope.  This choice of profiles corresponds to $\alpha \sim 2$. From the N-body simulations described above, we find that our synchronized halos have relative velocities of $\sim 20 ~ \rm{km/s}$ at small ($\sim  \rm{few}\times ~0.1~  \rm{kpc}$) separation.  For $\alpha =2$, Eq. \ref{ram} implies the core will be stable to a radius of $r=0.14r_{\rm vir}=0.13 \rm ~{kpc}$.  Even for $\alpha =1$, which is smaller than expected for realistic profiles, the gas cores will be stable to down to an orbital radius of $r=0.2r_{\rm vir}=0.18 \rm~{kpc}$.  Thus, while this should be verified in future hydrodynamical simulations, we do not expect ram pressure stripping to be a problem for the synchronized atomic cooling pairs scenario.  

We also investigate whether the gas cores in atomic cooling halos could be disrupted by tidal forces.  Tidal disruption is unimportant provided that the tidal force is smaller than the gravitational binding force at the surface of the core
\begin{equation}
\frac{G {M_{\rm tot}(R_{\rm core})}}{R_{\rm core}^2} >  \frac{G {M'_{\rm tot}(r-R_{\rm core})}}{(r - R_{\rm core})^2} - \frac{G {M'_{\rm tot}(r)}}{r^2}.
\end{equation}
Here the primed and unprimed masses denote different halos in the pair.
Dark matter dominates the surface gravity of the core before collapse, so we use the NFW profile to approximate $M_{\rm tot}$.  Checking with this approximation, we find that even if we assume there is a mass ratio of a few between our pair of halos, tidal disruption will not be important even near the radius where the cores begin to overlap.

\subsection{Photoevaporation}
As discussed above, if the gas in the progenitors of our atomic cooling halos is ionized and photoevaporated, DCBH formation could be prevented.  This is because the gas core would not be present as the halo crosses the atomic cooling threshold.

\begin{figure} 
\includegraphics[width=88mm]{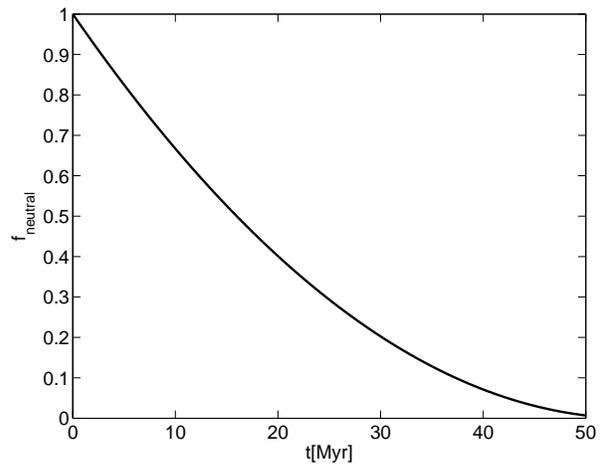}
\caption{Neutral gas fraction as a function of time for a  $2 \times 10^7 M_\odot$ halo at $z=10$ exposed to $J_{21}\sim 1000$ (assuming one ionizing photon per LW photon and a $T=10^5 ~\rm{K}$ blackbody spectrum) computed with the fits of \citet{2005MNRAS.361..405I}.  \label{photo} }
\end{figure}

We consider how long a halo just below the atomic cooling threshold ($2 \times 10^7 M_\odot$) can be exposed to ionizing radiation before being photoevaporated.  In Figure \ref{photo}, we plot the neutral fraction as a function of time for this halo using the fits from \cite{2005MNRAS.361..405I}, assuming $J_{21} = 1000$ and one ionizing photon per LW photon. This assumption is an overestimate for small pop III stars, but fairly accurate for large ones. The ratio of ionizing to LW photons expected from the calculations of \cite{2002A&A...382...28S} is $0.029/0.74/0.81/0.85/0.90$ for $5/15/25/40/200 ~ M_\odot$ stars.  In the $\sim 10  ~ \rm{Myr}$ it takes for a DCBH to form, most of the gas in the halo will remain neutral.  Thus, provided that our halos do not sit in a large-scale ionized region, we do not expect photoionization to be a problem.  This is mainly due to the fact that the synchronization of the star-forming and DCBH-forming halos ensures that the DCBH-forming halo is only exposed to ionizing radiation for a short time.

\subsection{Metal enrichment}
In addition to preventing photoevaporation, synchronization of atomic cooling halos can prevent metal contamination in the DCBH-forming halo. The lifetime of $5/15/25/40/200 ~ M_\odot$ pop III stars are $62/10/6.5/3.9/2.2 ~ \rm{Myr}$ respectively \citep{2002A&A...382...28S}. This suggests that given tight synchronization, a DCBH could potentially form before the stars in the partner halo die and eject their metals through supernova winds.  Additionally, a $100 ~\rm{km/s}$ wind would take $\sim5$ Myr, to travel $0.5~\rm{kpc}$.  Even if the metals do reach the DCBH before it is formed, the gas core forming a DCBH may have already collapsed to such high densities that metals from winds cannot penetrate.

\section{Abundance of Synchronized Pairs}
In this section, we estimate the abundance of DCBHs formed through the scenario of synchronized atomic cooling haloes described above.  We focus on redshifts near $z\sim 10$, since this is roughly the latest cosmic time that a DCBH can form and still grow as large as the black holes residing in the brightest $z \sim 6$ quasars (assuming 10 per cent radiative efficiency and Eddington limited accretion).  

We assume that one of the atomic cooling halos forms stars $t_{\rm coll} \sim10 ~\rm{Myr}$ after it reaches the cooling threshold ($M_{\rm c} = 3 \times 10^7 M_\odot$).  The DCBH-forming halo merges with this halo and becomes a subhalo near this time.   This subhalo must  cross the atomic cooling threshold during a window $\Delta t_{\rm sync}$ after stars are formed in its partner halo.  The size of this window will be set by how long the subhalo can be exposed to ionizing radiation without its gas core being photoevaporated.  A small $\Delta t_{\rm sync}$ may also allow a DCBH to form before it is polluted with metals from supernovae winds emitted by the nearby galaxy.  When the DCBH-forming subhalo crosses the atomic cooling threshold it must orbit its neighbor closely enough to be exposed to $J_{\rm crit}$, but far enough away to prevent ram pressure stripping.  The subhalo must remain in this distance range until the DCBH has formed $t_{\rm coll }\sim 10 ~\rm{Myr}$ later.  We refer to this range as the ``required orbital range.''  In the following subsections, we perform an analytic estimate for the number of DCBHs formed in this manner and compare to the number found directly in N-body simulations.

\subsection{N-body simulations}
To determine the abundance and properties of synchronized atomic cooling halos, we ran a set of N-body simulations with the publicly available code \textsc{gadget2} \citep{2005MNRAS.364.1105S}.  We performed 5 runs with different initial conditions created with \textsc{2lpt} \citep{2006MNRAS.373..369C}.  Our simulations each have $768^3$ particles in a $15 ~ \rm{cMpc}$ box.  This corresponds to $\sim 100$ particles per atomic cooling halo, which is enough to reliably track these halos as substructure of larger systems \citep{2012MNRAS.423.1200O}.  The simulations were started at $z=200$ and approximately 20 snapshots for each run were saved between $z=12$ and $z=10$ (with 3.33 Myr separation at the redshifts where we search for DCBH-forming pairs).  We used the \textsc{rockstar} halo finder \citep{2013ApJ...762..109B} to locate and track halos and subhalos over time.

\subsection{Analytic estimate}
The number of DCBHs formed per unit redshift per volume can be approximated by
\begin{equation}
\label{abund}
\frac{dn_{\rm DCBH}}{dz} \sim \frac{dn_{\rm cool}}{dz}  \left(  \frac{dn_{\rm cool}}{dz}\Delta z_{\rm sync}   \int^{\rm R.O.R.} dr 4\pi r^2[1+\xi(r)]  f_{\rm s}(r) \right ) ,
\end{equation}
where $\frac{dn_{\rm cool}}{dz}$ is the number density of halos which cross the cooling threshold between $z$ and $z+dz$, $\xi(r)$ is the two-point function which describes the enhancement of halo pairs due to clustering, and $\Delta z_{\rm sync} $  is the redshift range corresponding to $\Delta t_{\rm sync}$.  The fraction of sub-halos that are found at radius $r$ when they cross the cooling threshold and stay within the required orbital range until a DCBH is formed is denoted by $f_{\rm s}(r)$.  The quantity in parentheses represents the fraction of atomic cooling halos that have a synchronized partner which forms a DCBH.  The integral is evaluated over the required orbital range.

We examine our N-body simulations at $z \sim 10-11$ to determine values of the terms in Eq. \ref{abund}.
We find $\frac{dn_{\rm cool}}{dz} \sim 4 ~ \rm{cMpc}^{-3}$.  To estimate  $f_{\rm s}$, we identify all subhalos with $M/M_{\rm c} = 1 - 1.5$  in parent halos $M/M_{\rm c} =2  - 3 $ (such that the combined mass is approximately twice the atomic cooling mass).  We then determine the fraction that stay within the required orbital range for $t_{\rm coll}$.  We specifically identify subhalos because all pairs at the required orbital ranges we consider are in the same distinct halos.  If we were considering larger separations we would consider halo-halo pairs as well (i.e. not just subhalos).  We find that $f_{\rm s} \sim 0.2$ for a required orbital range of $r = 0.2-0.5 ~ \rm{kpc}$.  Due to the photoevaporation constraints described above, we assume the DCBH-forming halo has a 10 Myr window to start forming before it is photoevaporated, corresponding to $\Delta z_{\rm sync} \sim 0.16$ (assuming $J_{\rm crit}\sim 1000$, after $\Delta t_{\rm sync}+t_{\rm coll} = 20$ Myr, more than 40 per cent of the gas will remain neutral and we expect the core to still be intact).

\begin{figure} 
\includegraphics[width=88mm]{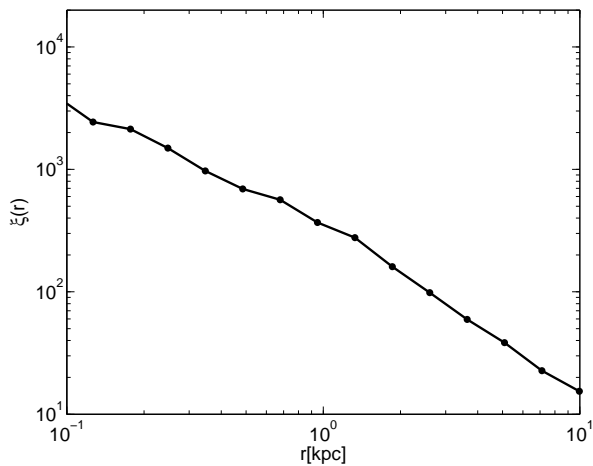}
\caption{Two-point correlation function, $\xi(M_1,M_2,z,r)$, at $z\sim10$ for mass bins $M_1/M_{\rm c} = 1.7 -2.3$ and $M_2/M_{\rm c} =  0.8-1.2$.  We use double the mass for the first bin because for our halo and subhalo pairs, the sum of both halo masses are included in the distinct (non-subhalo) halo.  In the first bin we do not include subhalos.  In the second bin, we consider halos and subhalos (with parent halos smaller than $M=2.3 M_{\rm c}$) that increase in mass by the next simulation snapshot.  At $r<1~\rm{kpc}$, we find that the correlation function is well approximated by the power law $\xi(r) = 700 \times (r/0.5 \rm{kpc})^{-1}$.
\label{two_pt} }
\end{figure}

We measure $\xi(r)$ directly from our N-body simulation.  When computing $\xi(r)$, we consider the correlation between two mass bins.  In the first bin, we consider distinct halos (i.e. not subhalos) with $M_1/M_{\rm c} = 1.7 -2.3$.  In the second bin, we consider all distinct halos and subhalos with parent halos smaller than $2.3 M_{\rm c}$, within a mass range of $M_2/M_{\rm c} = 0.8 -1.2$.  We additionally require the halos in the second bin to increase in mass by the next snapshot to make sure they could cross the cooling threshold.  In Figure \ref{two_pt}, we plot $\xi(r)$.  We find that $\xi(r)$ is described roughly by a power law for $r<1.0~\rm{kpc}$ , $\xi(r) = 700 \times (r/0.5 ~\rm{kpc})^{-1}$. 

Putting all of this together, we find that at $z \sim 10$,
\begin{equation}
\frac{dn_{\rm DCBH}}{dz} \sim  0.0003 \left (\frac{\Delta z_{\rm sync}}{0.16} \right ) ~ \rm{cMpc}^{-3} ,
\label{rate}
\end{equation}
assuming a required orbital range of $r = 0.2-0.5~ \rm{kpc}$.  We also estimate the abundance for orbital ranges of $r = 0.2-0.75 ~\rm{kpc}$ (corresponding to $f_s \sim 0.4$) and $r = 0.2-1.0~ \rm{kpc}$ (corresponding to $f_s \sim 0.5 $) and find $\frac{dn_{\rm DCBH}}{dz} \sim  0.0015 \left (\frac{\Delta z_{\rm sync}}{0.16} \right ) ~ \rm{cMpc}^{-3}$ and $\frac{dn_{\rm DCBH}}{dz} \sim  0.0036 \left (\frac{\Delta z_{\rm sync}}{0.16} \right ) ~ \rm{cMpc}^{-3}$  respectively. We note that the number density estimated in this section could be reduced due to metal enrichment from our halo pairs' progenitors. In \S4, we argue that depending on the LW background, self-enrichment from progenitors may be a small effect.

\subsection{Numerical estimate}
\begin{figure*} 
\includegraphics[width=88mm]{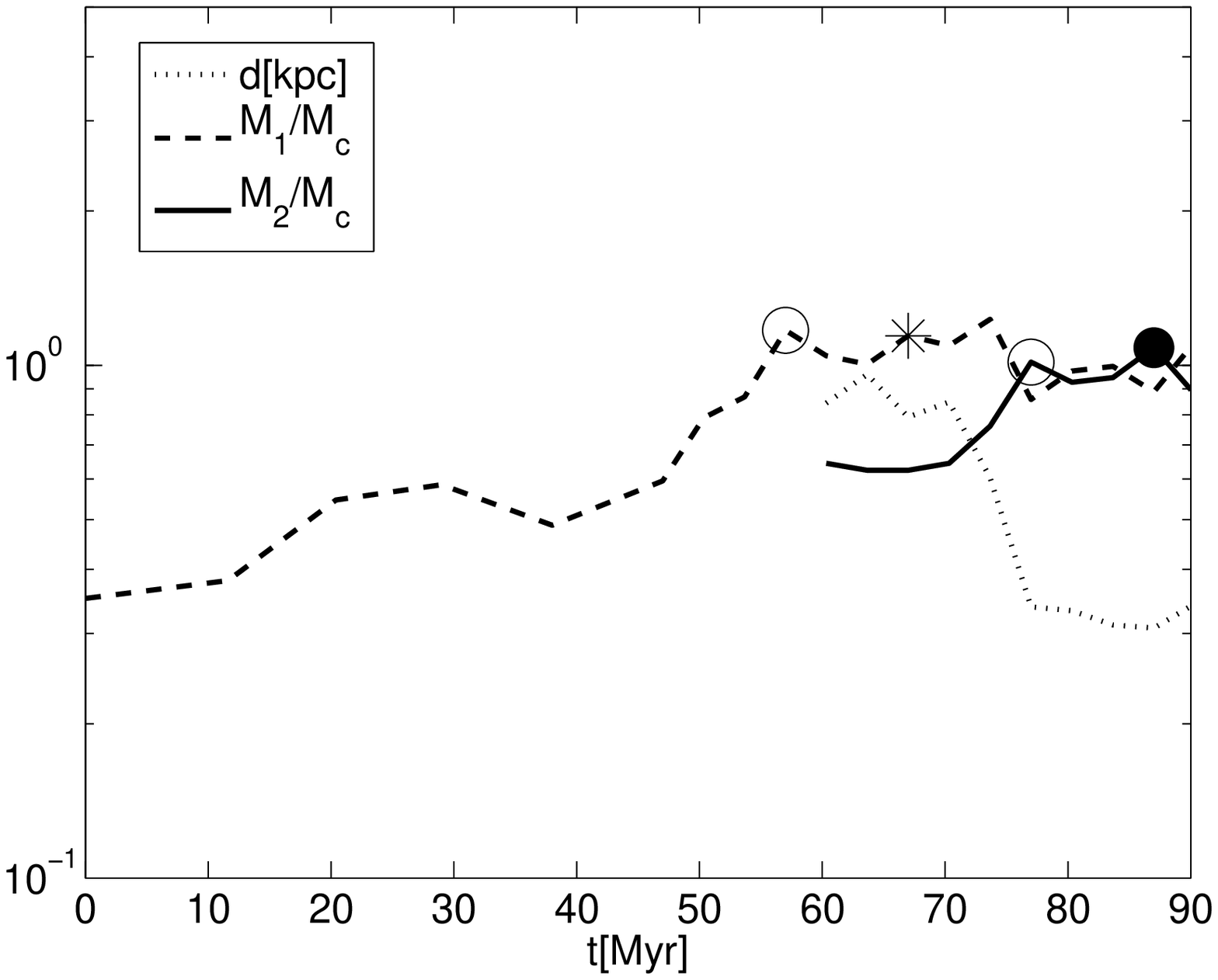}
\includegraphics[width=88mm]{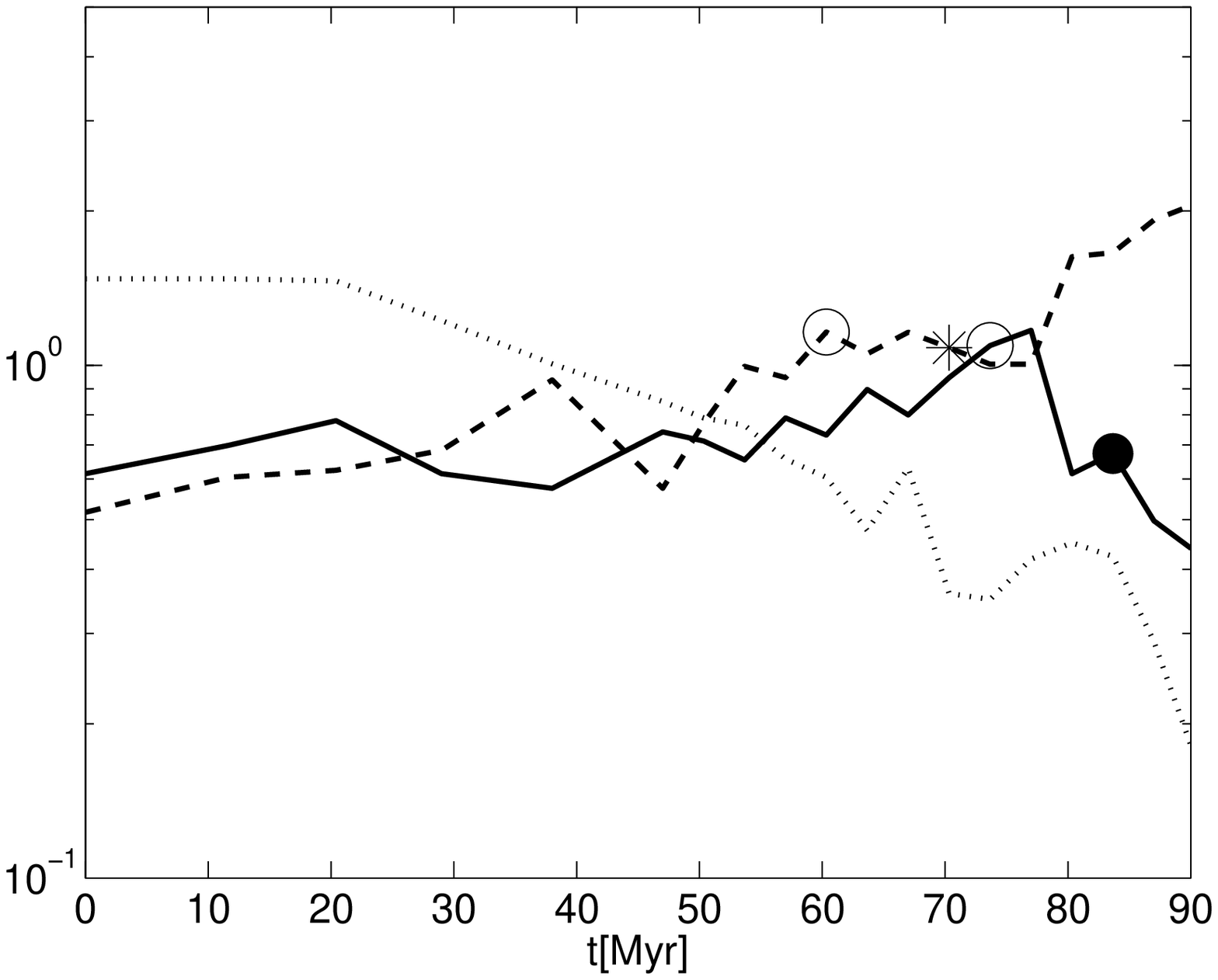}
\caption{The mass growth and separation of the synchronized pairs found in our N-body simulations for a required orbital range of $0.2-0.5 ~ \rm{kpc}$.  The dashed and solid curves show the mass divided by the atomic cooling threshold of the star-forming halo and DCBH-forming subhalo, respectively.  For the star-forming halos, we have subtracted the mass of the DCBH-forming subhalo which is initially included in the total mass by the halo finder.  The dotted curve shows the distance between the centers (i.e. density peaks) of these halos.  The open circles correspond to the time when the halos cross the atomic cooling threshold, the $*$'s indicate when stars are formed, and the large dark circles indicate the formation of DCBHs.  For the left panel, the DCBH-forming halo is not located by the halo finder until $t=60 ~{\rm Myr}$ due to its small number of particles, but is reliably tracked at $t>60 ~{\rm Myr}$.   \label{example} }
\end{figure*}

We search our N-body simulations for pairs of halos where the first halo crosses the cooling threshold and the second crosses it as a subhalo 10-20 Myr later.  When considering the mass of the first halo to reach the threshold, we do not include the mass from its subhalo partner.  After the second halo grows to the cooling mass, we check that the pair maintains a distance corresponding to the required orbital zone for 10 Myr.  This ensures they are not destroyed from ram pressure stripping yet still maintain $J_{\rm crit}$.  We search through $\Delta z \sim 0.25$ for three different values of the required orbital range: $0.2-0.5~ \rm{kpc}$, $0.2-0.75~ \rm{kpc}$, and $0.2-1.0 ~ \rm{kpc}$.  Based on the box size and our five simulations, the analytic estimates in the previous subsection predict that we will find 1.25, 6.2, and 15 synchronized pairs respectively for these required orbital ranges. In the N-body simulations we actually find 2, 5, and 17, which is in good agreement with our analytic expression. We plot the mass growth and separation of the pairs for the strictest required orbital range in Figure \ref{example}.  As we explain in the following section, the abundance of synchronized pairs we find is enough to explain the number density of the brightest $z=6$ quasars.

\section{Formation of high-redshift quasars}
For DCBHs formed through synchronized atomic cooling halos to grow into the brightest high-redshift quasars, they must end up in $\sim 10^{12} M_\odot$ dark matter halos by $z=6$.  The regions that eventually become $10^{12} M_\odot$ halos by $z=6$ are large-scale peaks in the density field where halos will form at earlier cosmic times.  This will lead to a locally higher redshift of reionization, which one might expect to photoevaporate the progenitors of atomic cooling halos, preventing DCBH formation.  However, here we argue that this will not prevent DCBH formation.  Instead it will shift their formation to higher redshift and potentially increase the DCBH abundance per comoving volume estimated above.

At $z=10$, the redshift we focus on above, the IGM is still expected to be mostly neutral \citep[see e.g.][]{2014MNRAS.439..725I}.  At higher redshift, the density of the IGM will be greater leading to an increased rate of recombination.  Thus, for a large-scale overdense region to reionize sooner, it must have a higher comoving density of ionizing photon sources (most likely galaxies in atomic cooling halos).  Additionally, the clustering of haloes in an overdense region will be enhanced, since halos form near the peak of a large-wavelength $k$-mode of the density field.  Because the number density will be increased and the clustering enhanced, we expect that the formation rate per comoving volume of DCBHs will be larger relative to our estimate at $z=10$ (see Eq. \ref{abund}).

The abundance of quasars hosting $\sim \rm{a ~few} \times 10^9 M_\odot$ SMBHs at $z=6$ is $\sim 1  ~ \rm{cGpc}^{-3}$ \citep{2006NewAR..50..665F}.  In $1 ~ \rm{cGpc}^3$ at $z=6$ there are $\sim 5500$ halos larger than $10^{12} M_\odot$ \citep{1999MNRAS.308..119S}.  The comoving volume in the regions that become these halos corresponds to $\sim 60$ times the volume of the $15 ~ \rm{cMpc}$ simulation box we use.  Thus, each $\rm{cGpc}^3$ region has roughly the equivalent of 60 of our boxes to make a DCBH.  Since the DCBH number density in these regions should be higher than what we calculate, the abundance of DCBHs created through synchronized atomic cooling halos could be high enough to explain the observed $z=6$ quasars.  This abundance suggests that we can require an even tighter synchronization window than the fiducial value used above.  For $\Delta t_{\rm sync} \sim 0.2 ~ \rm{Myr}$, there would likely still be enough DCBHs formed over $\Delta z\sim 1$ to explain the $z=6$ observations. This very close synchronization could enable the DCBH to form before stars in the neighboring halo die and expel supernovae winds.  Note that if we require the $z=6$ quasars to be in halos much larger than $10^{12}
~M_\odot$, the comoving volume where DCBHs can form and still end up in these halos goes down significantly due to their small number density.

An additional concern at higher redshifts is that the LW background could potentially fall low enough such that star formation occurs in the minihalo progenitors of our pairs of atomic cooling halos, leading to metal pollution and preventing DCBH formation.  To investigate this possibility, we estimate the change in the redshift of reionization for a region that becomes a $10^{12} M_\odot$ halo at $z=6$ using a simple model based on the extended Press-Schechter formalism \citep{1991ApJ...379..440B}.  We compute the mass-averaged ionization fraction, $Q$, as a function of time with the following differential equation
\begin{equation}
\frac{dQ}{dt} = N_{\rm ion}  \frac{dF(M_{\rm min})}{dt} - \alpha_{\rm B}  C n_{\rm H}Q,
\end{equation}
where $N_{\rm ion}$ is the number of ionizing photons released into the IGM per hydrogen atom collapsing into a dark matter halo and $F(M_{\rm min})$ is the fraction of mass that has collapsed into halos larger than $T_{\rm vir} = 10^4 \rm{K}$, which depends on both the overdensity and size of the region.  The second term on the right hand side is the recombination rate; $\alpha_{\rm B}$ is the case B recombination coefficient, $n_{\rm H}$ is the IGM hydrogen number density, and $C = \langle n_{\rm H}^2 \rangle/\langle n_{\rm H} \rangle^2$ is the IGM clumping parameter.  We find that for $C=3$ and $N_{\rm ion} = 10$, we obtain an ionization history similar to the simulations of \cite{2014MNRAS.439..725I} when we apply this estimate to a large region with mean cosmic density.  For a region that becomes a $10^{12} M_\odot$ halo at $z=6$ we find that reionization happens roughly $\Delta z=5$ earlier. Thus, we expect that DCBHs could form as late as $z\sim15$ in the regions that become the largest $z\sim6$ quasars before photoevaporation caused by reionization could pose significant problems.

\begin{figure} 
\includegraphics[width=88mm]{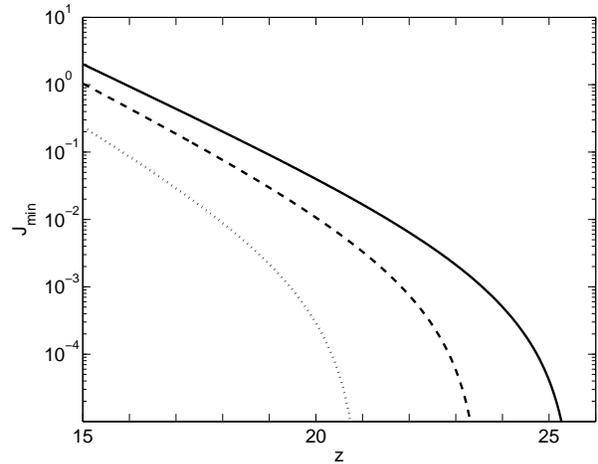}
\caption{  The minimum value of the LW background (in units of $10^{-21}~{\rm ergs ~s^{-1}cm^{-2}Hz^{-1}Sr^{-1}}$) required to shut off all star formation in the progenitors of a typical atomic cooling halo forming at $z=15$. The solid, dashed, and dotted lines are for regions with a baryon-dark matter streaming velocity equal to 0, 1, and 2 times the root-mean-square (RMS) value, respectively. The percentage of large-scale regions with streaming velocity greater than the RMS value is $39 \%$ ($0.74 \%$ for 2 times the RMS value). \label{minJ} }
\end{figure}

Next, we determine the required intensity of the LW background required to prevent star formation in the progenitors of an atomic cooling halo that forms a DCBH at $z=15$. We estimate the redshift evolution of the most-massive progenitor with eqn. 7 from \cite{2014MNRAS.440...50M}. There will be scatter for different merger histories, but this formula gives the typical mass as a function of redshift. Given the halo mass, we can determine the LW background required to shut off all star formation using the equations from \cite{2013MNRAS.432.2909F,2012MNRAS.424.1335F}. We point out that the baryon-dark matter streaming velocity \citep{2010PhRvD..82h3520T} can help to prevent star formation in minihalos. We use the ``optimal fit'' to cosmological simulations from \cite{2012MNRAS.424.1335F} to determine the impact of the streaming velocity on the minimum halo mass required to host star formation. In Figure \ref{minJ}, we plot the minimum intensity of the LW background necessary to shut off all star formation in minihalos for different values of the streaming velocity. 

The minimum values of the LW background required are lower than the predictions of \cite{2013MNRAS.432.2909F}, even in the ``saturated feedback'' case without star formation in minihalos. Additionally, since our DCBHs forming $z=6$ quasars are expected to be found in overdense regions, the local value of the LW background could be considerably higher than the mean \citep{2009ApJ...695.1430A}. Thus, metal self-enrichment may not significantly reduce the abundance of DCBHs estimated above. We note that a star formation efficiency of 10 per cent was assumed in \cite{2013MNRAS.432.2909F}. If the true value is much lower, star formation may occur in the progenitors of atomic cooling halos and pollute them with metals preventing DCBH formation. However, if only a small number of stars are formed, it is possible that all of these stars fall in the mass range leading to direct collapse into black holes without supernovae ($40-100 ~M_\odot$). This would prevent metal pollution and still permit DCBH formation. Since the fraction of halos that would remain pristine in this way depends on the LW background evolution and the IMF of pop III stars in minihalos, both of which are highly uncertain, we leave detailed estimates for future work. 

Additionally, we point out that the collapse time computed above scales as $t_{\rm coll} \propto (1+z)^{-3/2}$ due to the increase in density with redshift.  This could potentially make it easier to form DCBHs as there would be less time for photoevaporation and metal enrichment.  The increase in density of the gas cores of halos should also make the cores more self-shielding to ionizing radiation and metal pollution.  We also note that the halo mass corresponding to $10^4 \rm{K}$ scales as $M \propto (1+z)^{-3/2}$.  Thus a galaxy forming in an atomic cooling halo at $z=15$ could be nearly a factor of two smaller than we estimate in \S2.  However, the difference in luminosity associated with this mass change is much smaller than the uncertainty related to the pop III IMF.

\section{Discussion and Conclusions}
We have proposed a new scenario for the formation of DCBHs based on synchronized pairs of atomic cooling halos at small separation.  For such a pair, stars form in the first halo to reach the cooling threshold and produce the required LW flux to suppress molecular hydrogen cooling in the second halo, leading to the formation of a DCBH.  The second, DCBH-forming halo must cross the cooling threshold in a window $\Delta t_{\rm sync}$ after these stars are formed.  The size of this window is set by how quickly the gas core in the DCBH-forming halo is photoevaporated or polluted by metals from supernovae winds produced in the other halo.  To form a DCBH we assume that the LW intensity must remain above $J_{\rm crit}$ for $t_{\rm coll}$ after crossing the cooling threshold.  This requires the halos to remain within a distance range close enough to maintain $J_{\rm crit}$, but far enough to prevent significant ram pressure stripping of the DCBH-forming gas core.  We note that our assumption requiring the DCBH-forming halo to be exposed to $J_{\rm crit}$ for the entire $t_{\rm coll}$ is conservative, as a smaller background will be necessary in the early stages of collapse while the density remains relatively low.

We use a set of N-body simulations to estimate the abundance of DCBHs formed in this scenario.  We compare this estimate to an analytic expression for the abundance of DCBH and find good agreement for three different choices of the required orbital range.  The number density of synchronized pairs found in eqn. \ref{rate} for a required orbital range of $0.2 - 0.5~ \rm{kpc}$ and $\Delta t_{\rm sync} \sim 10~ \rm{Myr}$ is $\frac{dn_{\rm DCBH}}{dz} \sim 0.0003 ~ \rm{cMpc^{-3}}$ at $z\sim10$.  In \S4, we point out that the large dark matter halos that host the brightest $z\sim6$ quasars form from large-scale matter overdensities.  In these regions, star formation and reionization will be shifted to higher redshift, however we argue that there will be enough DCBHs formed to explain $z\sim6$ quasar observations, even with a tight ($\Delta t_{\rm sync} \sim 0.2~ {\rm Myr}$) synchronization window.  We appeal to the analytic arguments presented in \S4 (rather than directly checking in N-body simulations) because regions that form $10^{12} ~ M_\odot$ halos by $z=6$ are very rare.  Studying a statistical sample of these regions will require simulating a much larger box and resimulating the overdense regions at higher resolution.  We leave this for future work.  In \S4, we also compute the strength of the LW background required to prevent star formation and subsequent metal pollution in the progenitors of atomic cooling halos.

An advantage of the synchronized halo picture presented here over previous models of close pairs of halos  \citep[e.g.][]{2008MNRAS.391.1961D,2014arXiv1405.6743D} is that synchronization may circumvent the obstacles of photoevaporation and metal contamination from supernovae winds.  As explained in \S1, the progenitors of a large galaxy producing the critical LW intensity is likely to have photoevaporated the progenitors of nearby atomic cooling halos. Provided that the pair of halos sit in a large-scale region that has not yet been ionized, synchronization solves this problem. The DCBH-forming halo is only exposed to ionizing radiation for a short period.  Metal pollution from supernova winds could also be prevented by synchronization.  In the high-density regions that eventually become large quasars, DCBHs likely form at $z > 15$.  At high redshift, the increased density leads to a smaller $t_{\rm coll}$.  This, coupled with the fact that $J_{\rm crit}$ is likely only necessary near the end of the collapse, suggests that a DCBH may only require a strong LW background for a few Myr.  The lifetime of $40~ M_\odot$ metal free stars is $\sim 3.9 ~\rm{Myr}$.  Non-rotating stars between $40-100 ~M_\odot$ are expected to collapse directly into black holes without supernovae \citep{2003ApJ...591..288H}.  Thus, a DCBH could form before stars (smaller than $40~M_\odot$) die and eject material through supernovae. Even if the DCBH was not fully formed by the time these winds were ejected, the collapsing gas core may have reached a point with such high density that metals cannot penetrate \citep{2008ApJ...674..644C}.

The largest uncertainty associated with this scenario of DCBH formation is the IMF and star formation efficiency of metal free stars forming in atomic cooling halos.  As mentioned in \S2, if stars are mostly small ($< 5 ~M_\odot$), it may be more difficult to achieve $J_{\rm crit}$.  On the other hand, the short lifetime of more massive stars makes metal contamination more likely since there is less time for a DCBH to form before supernovae winds are ejected from the LW producing galaxy.  While DCBH formation may be possible for a wide range of possible IMFs, it seems that those with significant numbers of more massive stars will be most favorable.  The details of the IMF and star formation efficiency will determine the precise values of the required orbital range and $\Delta t_{\rm sync}$.   

We note that our simulations included dark matter only.  Future work utilizing hydrodynamical cosmological simulations with atomic cooling will be important for better constraining the abundance of DCBHs produced in this scenario.  As mentioned above, it will be beneficial for these simulations to zoom-in on large-scale overdense regions that become $\sim 10^{12} M_\odot$ halos at $z\sim6$ to see the possible DCBH density enhancement discussed in \S4.  

Overall, we consider the synchronized atomic cooling scenario an attractive way to explain the brightest $z=6$ quasars.  Similar channels of DCBH formation without synchronization may not be able to produce the required number density of DCBHs when metal enrichment and photoevaporation are taken into account.

\section*{Acknowledgements}
We thank Kohei Inayoshi, Jemma Wolcott-Green, and Mark Dijkstra for useful discussions.  EV was supported by the Columbia Prize Postdoctoral Fellowship in the Natural Sciences.  ZH was supported by NASA grant NNX11AE05G.  GLB was supported by National Science Foundation grant 1008134 and NASA grant NNX12AH41G.  This work used the Extreme Science and Engineering Discovery Environment (XSEDE), which is supported by National Science Foundation grant number ACI-1053575.  The numerical simulations utilized in this study were carried out at the Texas Advanced Computing Center.

\bibliography{pairs_paper}

\begin{thebibliography}{}

\bibitem[\protect\citeauthoryear{{Ade}, {Aghanim}, {Armitage-Caplan}, {Arnaud},
  {Ashdown}, {Atrio-Barandela}, {Aumont}, {Baccigalupi}, {Banday} \& et
  al.}{{Ade} et~al.}{2013}]{2013arXiv1303.5076P}
{Ade} P.~A.~R.,  {Aghanim} N.,  {Armitage-Caplan} C.,  {Arnaud} M.,  {Ashdown}
  M.,  {Atrio-Barandela} F.,  {Aumont} J.,  {Baccigalupi} C.,  {Banday} A.~J.,
    et al. 2013, astro-ph: 1303.5076

\bibitem[\protect\citeauthoryear{{Agarwal}, {Dalla Vecchia}, {Johnson},
  {Khochfar} \& {Paardekooper}}{{Agarwal} et~al.}{2014}]{2014arXiv1403.5267A}
{Agarwal} B.,  {Dalla Vecchia} C.,  {Johnson} J.~L.,  {Khochfar} S.,
  {Paardekooper} J.-P.,  2014, astro-ph: 1403.5267

\bibitem[\protect\citeauthoryear{{Agarwal}, {Khochfar}, {Johnson}, {Neistein},
  {Dalla Vecchia} \& {Livio}}{{Agarwal} et~al.}{2012}]{2012MNRAS.425.2854A}
{Agarwal} B.,  {Khochfar} S.,  {Johnson} J.~L.,  {Neistein} E.,  {Dalla
  Vecchia} C.,    {Livio} M.,  2012, \mnras, 425, 2854

\bibitem[\protect\citeauthoryear{{Ahn}, {Shapiro}, {Iliev}, {Mellema} \&
  {Pen}}{{Ahn} et~al.}{2009}]{2009ApJ...695.1430A}
{Ahn} K.,  {Shapiro} P.~R.,  {Iliev} I.~T.,  {Mellema} G.,    {Pen} U.-L.,
  2009, \apj, 695, 1430

\bibitem[\protect\citeauthoryear{{Alvarez}, {Wise} \& {Abel}}{{Alvarez}
  et~al.}{2009}]{2009ApJ...701L.133A}
{Alvarez} M.~A.,  {Wise} J.~H.,    {Abel} T.,  2009, \apjl, 701, L133

\bibitem[\protect\citeauthoryear{{Barkana} \& {Loeb}}{{Barkana} \&
  {Loeb}}{2004}]{2004ApJ...609..474B}
{Barkana} R.,  {Loeb} A.,  2004, \apj, 609, 474

\bibitem[\protect\citeauthoryear{{Behroozi}, {Wechsler} \& {Wu}}{{Behroozi}
  et~al.}{2013}]{2013ApJ...762..109B}
{Behroozi} P.~S.,  {Wechsler} R.~H.,    {Wu} H.-Y.,  2013, \apj, 762, 109

\bibitem[\protect\citeauthoryear{{Bond}, {Cole}, {Efstathiou} \&
  {Kaiser}}{{Bond} et~al.}{1991}]{1991ApJ...379..440B}
{Bond} J.~R.,  {Cole} S.,  {Efstathiou} G.,    {Kaiser} N.,  1991, \apj, 379,
  440

\bibitem[\protect\citeauthoryear{{Cen} \& {Riquelme}}{{Cen} \&
  {Riquelme}}{2008}]{2008ApJ...674..644C}
{Cen} R.,  {Riquelme} M.~A.,  2008, \apj, 674, 644

\bibitem[\protect\citeauthoryear{{Crocce}, {Pueblas} \& {Scoccimarro}}{{Crocce}
  et~al.}{2006}]{2006MNRAS.373..369C}
{Crocce} M.,  {Pueblas} S.,    {Scoccimarro} R.,  2006, \mnras, 373, 369

\bibitem[\protect\citeauthoryear{{Dijkstra}, {Ferrara} \&
  {Mesinger}}{{Dijkstra} et~al.}{2014}]{2014arXiv1405.6743D}
{Dijkstra} M.,  {Ferrara} A.,    {Mesinger} A.,  2014, astro-ph: 1405.6743

\bibitem[\protect\citeauthoryear{{Dijkstra}, {Haiman}, {Mesinger} \&
  {Wyithe}}{{Dijkstra} et~al.}{2008}]{2008MNRAS.391.1961D}
{Dijkstra} M.,  {Haiman} Z.,  {Mesinger} A.,    {Wyithe} J.~S.~B.,  2008,
  \mnras, 391, 1961

\bibitem[\protect\citeauthoryear{{Fan}}{{Fan}}{2006}]{2006NewAR..50..665F}
{Fan} X.,  2006, \nar, 50, 665

\bibitem[\protect\citeauthoryear{{Fialkov}, {Barkana}, {Tseliakhovich} \&
  {Hirata}}{{Fialkov} et~al.}{2012}]{2012MNRAS.424.1335F}
{Fialkov} A.,  {Barkana} R.,  {Tseliakhovich} D.,    {Hirata} C.~M.,  2012,
  \mnras, 424, 1335

\bibitem[\protect\citeauthoryear{{Fialkov}, {Barkana}, {Visbal},
  {Tseliakhovich} \& {Hirata}}{{Fialkov} et~al.}{2013}]{2013MNRAS.432.2909F}
{Fialkov} A.,  {Barkana} R.,  {Visbal} E.,  {Tseliakhovich} D.,    {Hirata}
  C.~M.,  2013, \mnras, 432, 2909

\bibitem[\protect\citeauthoryear{{Finlator}, {Oh}, {{\"O}zel} \&
  {Dav{\'e}}}{{Finlator} et~al.}{2012}]{2012MNRAS.427.2464F}
{Finlator} K.,  {Oh} S.~P.,  {{\"O}zel} F.,    {Dav{\'e}} R.,  2012, \mnras,
  427, 2464

\bibitem[\protect\citeauthoryear{{Haiman}}{{Haiman}}{2013}]{2013ASSL..396..293H}
{Haiman} Z.,  2013 Vol.~396 of Astrophysics and Space Science Library.
p.~293

\bibitem[\protect\citeauthoryear{{Heger}, {Fryer}, {Woosley}, {Langer} \&
  {Hartmann}}{{Heger} et~al.}{2003}]{2003ApJ...591..288H}
{Heger} A.,  {Fryer} C.~L.,  {Woosley} S.~E.,  {Langer} N.,    {Hartmann}
  D.~H.,  2003, \apj, 591, 288

\bibitem[\protect\citeauthoryear{{Iliev}, {Mellema}, {Ahn}, {Shapiro}, {Mao} \&
  {Pen}}{{Iliev} et~al.}{2014}]{2014MNRAS.439..725I}
{Iliev} I.~T.,  {Mellema} G.,  {Ahn} K.,  {Shapiro} P.~R.,  {Mao} Y.,    {Pen}
  U.-L.,  2014, \mnras, 439, 725

\bibitem[\protect\citeauthoryear{{Iliev}, {Shapiro} \& {Raga}}{{Iliev}
  et~al.}{2005}]{2005MNRAS.361..405I}
{Iliev} I.~T.,  {Shapiro} P.~R.,    {Raga} A.~C.,  2005, \mnras, 361, 405

\bibitem[\protect\citeauthoryear{{Johnson} \& {Bromm}}{{Johnson} \&
  {Bromm}}{2007}]{2007MNRAS.374.1557J}
{Johnson} J.~L.,  {Bromm} V.,  2007, \mnras, 374, 1557

\bibitem[\protect\citeauthoryear{{Johnson}, {Whalen}, {Agarwal}, {Paardekooper}
  \& {Khochfar}}{{Johnson} et~al.}{2014}]{2014arXiv1405.2081J}
{Johnson} J.~L.,  {Whalen} D.~J.,  {Agarwal} B.,  {Paardekooper} J.-P.,
  {Khochfar} S.,  2014, astro-ph: 1405.2081

\bibitem[\protect\citeauthoryear{{Latif}, {Bovino}, {Van Borm}, {Grassi},
  {Schleicher} \& {Spaans}}{{Latif} et~al.}{2014}]{2014arXiv1404.5773L}
{Latif} M.~A.,  {Bovino} S.,  {Van Borm} C.,  {Grassi} T.,  {Schleicher}
  D.~R.~G.,    {Spaans} M.,  2014, astro-ph: 1404.5773

\bibitem[\protect\citeauthoryear{{Leitherer}, {Schaerer}, {Goldader},
  {Delgado}, {Robert}, {Kune}, {de Mello}, {Devost} \& {Heckman}}{{Leitherer}
  et~al.}{1999}]{1999ApJS..123....3L}
{Leitherer} C.,  {Schaerer} D.,  {Goldader} J.~D.,  {Delgado} R.~M.~G.,
  {Robert} C.,  {Kune} D.~F.,  {de Mello} D.~F.,  {Devost} D.,    {Heckman}
  T.~M.,  1999, \apjs, 123, 3

\bibitem[\protect\citeauthoryear{{Madau}, {Haardt} \& {Dotti}}{{Madau}
  et~al.}{2014}]{2014ApJ...784L..38M}
{Madau} P.,  {Haardt} F.,    {Dotti} M.,  2014, \apjl, 784, L38

\bibitem[\protect\citeauthoryear{{McCarthy}, {Frenk}, {Font}, {Lacey}, {Bower},
  {Mitchell}, {Balogh} \& {Theuns}}{{McCarthy}
  et~al.}{2008}]{2008MNRAS.383..593M}
{McCarthy} I.~G.,  {Frenk} C.~S.,  {Font} A.~S.,  {Lacey} C.~G.,  {Bower}
  R.~G.,  {Mitchell} N.~L.,  {Balogh} M.~L.,    {Theuns} T.,  2008, \mnras,
  383, 593

\bibitem[\protect\citeauthoryear{{Milosavljevi{\'c}} \&
  {Bromm}}{{Milosavljevi{\'c}} \& {Bromm}}{2014}]{2014MNRAS.440...50M}
{Milosavljevi{\'c}} M.,  {Bromm} V.,  2014, \mnras, 440, 50

\bibitem[\protect\citeauthoryear{{Milosavljevi{\'c}}, {Bromm}, {Couch} \&
  {Oh}}{{Milosavljevi{\'c}} et~al.}{2009}]{2009ApJ...698..766M}
{Milosavljevi{\'c}} M.,  {Bromm} V.,  {Couch} S.~M.,    {Oh} S.~P.,  2009,
  \apj, 698, 766

\bibitem[\protect\citeauthoryear{{Navarro}, {Frenk} \& {White}}{{Navarro}
  et~al.}{1997}]{1997ApJ...490..493N}
{Navarro} J.~F.,  {Frenk} C.~S.,    {White} S.~D.~M.,  1997, \apj, 490, 493

\bibitem[\protect\citeauthoryear{{Onions}, {Knebe}, {Pearce}, {Muldrew}, {Lux},
  {Knollmann}, {Ascasibar}, {Behroozi}, {Elahi}, {Han}, {Maciejewski},
  {Merch{\'a}n}, {Neyrinck}, {Ruiz}, {Sgr{\'o}}, {Springel} \&
  {Tweed}}{{Onions} et~al.}{2012}]{2012MNRAS.423.1200O}
{Onions} J.,  {Knebe} A.,  {Pearce} F.~R.,  {Muldrew} S.~I.,  {Lux} H.,
  {Knollmann} S.~R.,  {Ascasibar} Y.,  {Behroozi} P.,  {Elahi} P.,  {Han} J.,
  {Maciejewski} M.,  {Merch{\'a}n} M.~E.,  {Neyrinck} M.,  {Ruiz} A.~N.,
  {Sgr{\'o}} M.~A.,  {Springel} V.,    {Tweed} D.,  2012, \mnras, 423, 1200

\bibitem[\protect\citeauthoryear{{Pawlik}, {Schaye} \& {van
  Scherpenzeel}}{{Pawlik} et~al.}{2009}]{2009MNRAS.394.1812P}
{Pawlik} A.~H.,  {Schaye} J.,    {van Scherpenzeel} E.,  2009, \mnras, 394,
  1812

\bibitem[\protect\citeauthoryear{{Schaerer}}{{Schaerer}}{2002}]{2002A&A...382...28S}
{Schaerer} D.,  2002, \aap, 382, 28

\bibitem[\protect\citeauthoryear{{Shang}, {Bryan} \& {Haiman}}{{Shang}
  et~al.}{2010}]{2010MNRAS.402.1249S}
{Shang} C.,  {Bryan} G.~L.,    {Haiman} Z.,  2010, \mnras, 402, 1249

\bibitem[\protect\citeauthoryear{{Sheth} \& {Tormen}}{{Sheth} \&
  {Tormen}}{1999}]{1999MNRAS.308..119S}
{Sheth} R.~K.,  {Tormen} G.,  1999, \mnras, 308, 119

\bibitem[\protect\citeauthoryear{{Springel}}{{Springel}}{2005}]{2005MNRAS.364.1105S}
{Springel} V.,  2005, \mnras, 364, 1105

\bibitem[\protect\citeauthoryear{{Tanaka} \& {Haiman}}{{Tanaka} \&
  {Haiman}}{2009}]{2009ApJ...696.1798T}
{Tanaka} T.,  {Haiman} Z.,  2009, \apj, 696, 1798

\bibitem[\protect\citeauthoryear{{Tanaka}}{{Tanaka}}{2014}]{2014arXiv1406.3023T}
{Tanaka} T.~L.,  2014, ArXiv e-prints

\bibitem[\protect\citeauthoryear{{Tseliakhovich} \& {Hirata}}{{Tseliakhovich}
  \& {Hirata}}{2010}]{2010PhRvD..82h3520T}
{Tseliakhovich} D.,  {Hirata} C.,  2010, \prd, 82, 083520

\bibitem[\protect\citeauthoryear{{Visbal}, {Haiman} \& {Bryan}}{{Visbal}
  et~al.}{2014}]{2014arXiv1403.1293V}
{Visbal} E.,  {Haiman} Z.,    {Bryan} G.~L.,  2014, \mnras, 442, L100

\bibitem[\protect\citeauthoryear{{Visbal}, {Haiman}, {Terrazas}, {Bryan} \&
  {Barkana}}{{Visbal} et~al.}{2014}]{2014arXiv1402.0882V}
{Visbal} E.,  {Haiman} Z.,  {Terrazas} B.,  {Bryan} G.~L.,    {Barkana} R.,
  2014, astro-ph: 1402.0882

\bibitem[\protect\citeauthoryear{{Volonteri}}{{Volonteri}}{2010}]{2010A&ARv..18..279V}
{Volonteri} M.,  2010, \aapr, 18, 279

\bibitem[\protect\citeauthoryear{{Wolcott-Green} \& {Haiman}}{{Wolcott-Green}
  \& {Haiman}}{2012}]{2012MNRAS.425L..51W}
{Wolcott-Green} J.,  {Haiman} Z.,  2012, \mnras, 425, L51

\bibitem[\protect\citeauthoryear{{Wolcott-Green}, {Haiman} \&
  {Bryan}}{{Wolcott-Green} et~al.}{2011}]{2011MNRAS.418..838W}
{Wolcott-Green} J.,  {Haiman} Z.,    {Bryan} G.~L.,  2011, \mnras, 418, 838

\bibitem[\protect\citeauthoryear{{Yue}, {Ferrara}, {Salvaterra}, {Xu} \&
  {Chen}}{{Yue} et~al.}{2014}]{2014MNRAS.440.1263Y}
{Yue} B.,  {Ferrara} A.,  {Salvaterra} R.,  {Xu} Y.,    {Chen} X.,  2014,
  \mnras, 440, 1263

\end{thebibliography}
\end{document}